# On the quantum polarization and entanglement of superposition of two two-mode coherent states

Joao Batista Rosa Silva and Rubens Viana Ramos

joaobrs@deti.ufc.br        rubens@deti.ufc.br

Department of Teleinformatic Engineering, Federal University of Ceará, 60455-760, C.P. 6007, Fortaleza-Ce, Brazil

## Abstract

This work discuss the entanglement and quantum polarization of superpositions of two-mode coherent states of the types $|\psi_1\rangle = N_1(|\alpha,\beta\rangle + |\beta,\alpha\rangle)$, $|\psi_2\rangle = N_2(|-\alpha,-\alpha\rangle + |\alpha,\alpha\rangle)$ and $|\psi_3\rangle = N_3(|\alpha,0\rangle + |0,\alpha\rangle)$. We use the concurrence to measure their entanglements and the quantum Stokes parameters and the $Q$ function in order to analyze their polarization and degree of polarization.

## 1. Introduction

Quantum polarization is an important property that has extensively being used in quantum information processing. Regarding coherent states, their quantum polarization has similarities with polarization of classical light: the average values of the Stokes parameters for coherent states coincide with the classical values of the Stokes parameters. On the other hand, the degree of polarization of classical light does not depend on its intensity, while the degree of polarization of coherent states decreases when the optical power decreases. Furthermore, the quantum Stokes parameters $\hat{S}_1, \hat{S}_2$ and $\hat{S}_3$ do not commute, that is, it is not possible to known the values of any two of them simultaneously without uncertainties. This fact has been used in [1] in order to create a continuous variable quantum key distribution system. Basically, polarization of coherent states has been used for quantum key distribution [1-4]. On the other hand, the use of superposition of coherent states for quantum computing purposes has been considered in [5-8]. However, from the best of our knowledge, the quantum polarization of superposition of coherent states has not been investigated. In this direction, here we use the quantum Stokes parameters and the $Q$ function in order to analyze the quantum polarization of superposition of coherent states of the type $|\psi\rangle = N(|\alpha,\beta\rangle + |\varepsilon,\lambda\rangle)$

(*N* is the normalization constant), as well we use the concurrence to measure their entanglements.

This work is outlined as follows: in Section 2 a review of quantum polarization is presented; in Section 3 the quantum polarization of superposition of coherent states is analyzed; in Section 4 the entanglement is calculated; Finally, Section 5 brings the conclusions.

## 2. Review of quantum polarization

The quantum Stokes operators are defined as [9-11]

$$\hat{S}_0 = \hat{a}_1^\dagger \hat{a}_1 + \hat{a}_2^\dagger \hat{a}_2, \tag{1}$$

$$\hat{S}_1 = \hat{a}_1^\dagger \hat{a}_1 - \hat{a}_2^\dagger \hat{a}_2, \tag{2}$$

$$\hat{S}_2 = \hat{a}_1^\dagger \hat{a}_2 + \hat{a}_2^\dagger \hat{a}_1, \tag{3}$$

$$\hat{S}_3 = i\left(\hat{a}_2^\dagger \hat{a}_1 - \hat{a}_1^\dagger \hat{a}_2\right), \tag{4}$$

$$\left[\hat{S}_j, \hat{S}_k\right] = i2\hat{S}_m, \quad \left[\hat{S}_0, \hat{S}_j\right] = 0 \text{ for } j,k,m \in \{1,2,3\}. \tag{5}$$

In (1)-(4) $\hat{a}_1^\dagger(\hat{a}_1)$ and $\hat{a}_2^\dagger(\hat{a}_2)$ are, respectively, the creation (annihilation) operators of the modes 1 and 2. The average values of the quantum Stokes parameters of a coherent state are equal to its classical Stokes parameters values. However, since there exist a variance on the Stokes parameters values, the polarization is not well defined. The mean values and the variances of the Stokes parameters of the two-mode coherent state $|\alpha,\beta\rangle$ are given by:

$$|\alpha,\beta\rangle = \sum_{n=0}^{\infty} e^{-\frac{|\alpha|^2}{2}} \frac{\alpha^n}{\sqrt{n!}} |n\rangle \otimes \sum_{k=0}^{\infty} e^{-\frac{|\beta|^2}{2}} \frac{\beta^k}{\sqrt{k!}} |k\rangle \tag{6}$$

$$\langle \hat{S}_1 \rangle = |\alpha|^2 - |\beta|^2, \quad \langle \hat{S}_1^2 \rangle = \left(|\alpha|^2 - |\beta|^2\right)^2 + |\alpha|^2 + |\beta|^2, \quad V_1 = |\alpha|^2 + |\beta|^2 \tag{7}$$

$$\langle \hat{S}_2 \rangle = \alpha^*\beta + \alpha\beta^*, \quad \langle \hat{S}_2^2 \rangle = \left(\alpha^*\beta\right)^2 + \left(\alpha\beta^*\right)^2 + |\alpha|^2 + |\beta|^2 + 2|\alpha|^2|\beta|^2, \quad V_2 = |\alpha|^2 + |\beta|^2 \tag{8}$$

$$\langle \hat{S}_3 \rangle = i\left(\alpha\beta^* - \alpha^*\beta\right), \quad \langle \hat{S}_3^2 \rangle = -\left(\alpha^*\beta\right)^2 - \left(\alpha\beta^*\right)^2 + |\alpha|^2 + |\beta|^2 + 2|\alpha|^2|\beta|^2, \quad V_3 = |\alpha|^2 + |\beta|^2 \tag{9}$$

$$V_i = \langle \hat{S}_i^2 \rangle - \langle \hat{S}_i \rangle^2, = \langle \alpha,\beta|\hat{S}_i^2|\alpha,\beta\rangle - \langle \alpha,\beta|\hat{S}_i|\alpha,\beta\rangle^2, \quad i \in \{0,1,2,3\} \text{ e } j \in \{1,2\}. \tag{10}$$

The larger is the optical power the larger are the variances of all three parameters.

In order to apply a phase shift $\phi$ between the horizontal and vertical modes of $|\alpha,\beta\rangle$, resulting in $|\alpha e^{i\phi/2}, \beta e^{-i\phi/2}\rangle$, the operator used is $C(\phi) = \exp(i\phi\hat{S}_1/2)$. On the other hand, if the goal is to apply a geometric rotation of $\theta$ in the polarization, then the operator to be used is $R(\theta) = \exp(i\theta \hat{S}_3)$. Thus, $R(\theta)|\alpha,\beta\rangle = |\beta \operatorname{sen}\theta + \alpha\cos\theta, \beta\cos\theta - \alpha\sin\theta\rangle$.

Classically, the light is considered unpolarized if its Stokes parameters vanish. Quantum-mechanically, this condition (in average) is required but it is not sufficient. In a more general way, a light beam can be considered unpolarized if its observables do not change after an application of a geometric rotation and/or a phase shift between the components. These conditions are mathematically described by [12]:

$$\left[\rho, \hat{S}_3\right] = \left[\rho, \hat{S}_1\right] = 0, \tag{11}$$

where $\rho$ is the density matrix of the light quantum state. The general form of the quantum state of an unpolarized light is [13-14]

$$\rho = \sum_n p_n \frac{1}{n+1} \sum_{k=0}^{n} |k\rangle|n-k\rangle\langle k|\langle n-k|, \tag{12}$$

where $p_n$ is the total photon number probability distribution.

Measures for the quantum polarization degree have been proposed [15,16] and here we consider the measure based on the $Q$ function [15]:

$$|n,\theta,\phi\rangle = \sum_{m=0}^{n} \binom{n}{m}^{1/2} \left[\operatorname{sen}\left(\frac{\theta}{2}\right)\right]^{n-m} \left[\cos\left(\frac{\theta}{2}\right)\right]^{m} e^{-im\phi} |m\rangle|n-m\rangle \tag{13}$$

$$Q(\theta,\phi) = \sum_{n=0}^{\infty} \frac{n+1}{4\pi} \langle n,\theta,\phi|\rho|n,\theta,\phi\rangle. \tag{14}$$

$$D = 4\pi \int_0^{2\pi} \int_0^{\pi} \left[Q(\theta,\phi) - \frac{1}{4\pi}\right]^2 \operatorname{sen}(\theta)\,d\theta d\phi, \tag{15}$$

$$P = \frac{D}{1+D}, \quad 0 \leq P \leq 1, \tag{16}$$

In (15) $1/(4\pi)$ is the $Q$ function of the unpolarized light whose quantum state is given by (12). For the two-mode coherent state $||\alpha|\exp(i\phi_\alpha),|\beta|\exp(i\phi_\beta)\rangle$, the $Q$ function is [17]

$$Q(\theta,\phi) = \frac{e^{-(|\alpha|^2+|\beta|^2)}}{4\pi}(1+z)e^z, \tag{17}$$

$$z = \left[|\alpha|\cos\left(\frac{\theta}{2}\right)\cos(\phi_\alpha+\phi) + |\beta|\operatorname{sen}\left(\frac{\theta}{2}\right)\cos(\phi_\beta)\right]^2 + \left[|\alpha|\cos\left(\frac{\theta}{2}\right)\operatorname{sen}(\phi_\alpha+\phi) + |\beta|\operatorname{sen}\left(\frac{\theta}{2}\right)\operatorname{sen}(\phi_\beta)\right]^2. \tag{18}$$

Using (13)-(16) the quantum degree of polarization of the state $|\alpha,0\rangle$ can be calculated analytically

$$P = 1 - \frac{4|\alpha|^2}{1+2|\alpha|^2\left(1+|\alpha|^2\right)-e^{-2|\alpha|^2}}. \qquad (19)$$

When $|\alpha|^2 \gg 1$, (19) can be approximated by $P \approx 1 - 2/|\alpha|^2$ showing that the larger the optical power the larger is the quantum degree of polarization. The quantum degree of polarization, of the states $|0,\pm\alpha\rangle$ ($|V\rangle$), $|\pm\alpha,0\rangle$ ($|H\rangle$), $|\pm\alpha,\mp\alpha\rangle$ ($|-\pi/4\rangle$) and $|\pm\alpha,\pm\alpha\rangle$ ($|\pi/4\rangle$) were obtained solving numerically (13)-(16) and they are shown in Fig. 1.

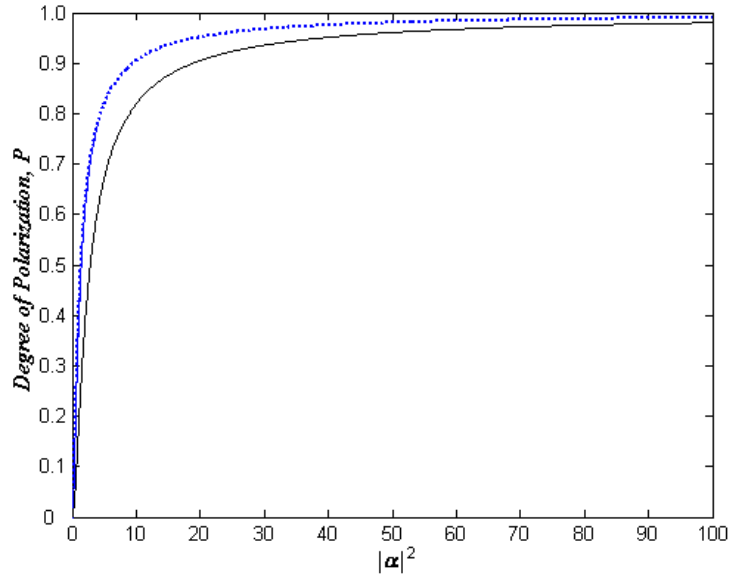

**Fig. 1: Quantum degree of polarization, *P*, versus $|\alpha|^2$ for the two-mode coherent states $\{|0,\alpha\rangle,|\alpha,0\rangle,|0,-\alpha\rangle,|-\alpha,0\rangle\}$ (line) and $\{|\alpha,\alpha\rangle,|-\alpha,-\alpha\rangle,|-\alpha,\alpha\rangle,|\alpha,-\alpha\rangle\}$ (dot).**

In Fig. 1, the diagonal states have a larger quantum degree of polarization, when compared to horizontal and vertical states, because they have a larger (total) mean photon number. As an illustration, using (13)-(14) one can obtain the *Q* functions for the states $|2,0\rangle$ (*H*), $|0,2\rangle$ (*V*), $|2,2\rangle$ ($\pi/4$), $|2,-2\rangle$ ($-\pi/4$), $|2,i2\rangle$ (*RC* – right circular) and $|2,-i2\rangle$ (*LC* – left circular), shown in Fig. 2. In this figure, $x=Q(\theta,\phi)\sin(\theta)\cos(\phi)$, $y=Q(\theta,\phi)\sin(\theta)\sin(\phi)$ and $z=Q(\theta,\phi)\cos(\theta)$.

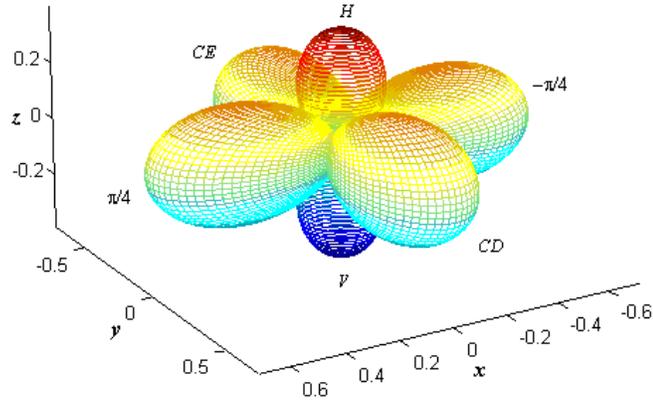

**Fig. 2 : Graphs of $Q$ fucntions of the states $|H\rangle=|2,0\rangle$, $|V\rangle=|0,2\rangle$, $|\pi/4\rangle=|2,2\rangle$, $|-\pi/4\rangle=|2,-2\rangle$, $|RC\rangle=|2,i2\rangle$ and $|LC\rangle=|2,-i2\rangle$.**

One can also analyze the polarization using the amplitude operators (similar to position operator of the harmonic oscillator) [15]:

$$\hat{x} = \frac{1}{2}\left(\hat{a}_x + \hat{a}_x^\dagger\right) \text{ e } \hat{y} = \frac{1}{2}\left(\hat{a}_y + \hat{a}_y^\dagger\right). \quad (20)$$

Considering $x$ and $y$ the eigenvalues corresponding, respectively, to the eigenstates $|x\rangle$ and $|y\rangle$, the amplitude probability distribution is given by

$$\mathcal{P} = \left|\langle x, y | \alpha, \beta \rangle\right|^2. \quad (21)$$

Using numerical simulation, we show the amplitude probabilities distributions for the quantum states $|2,0\rangle$ and $|0,2\rangle$ in Fig. 3, and $|2,2\rangle$ and $|2,-2\rangle$ in Fig. 4.

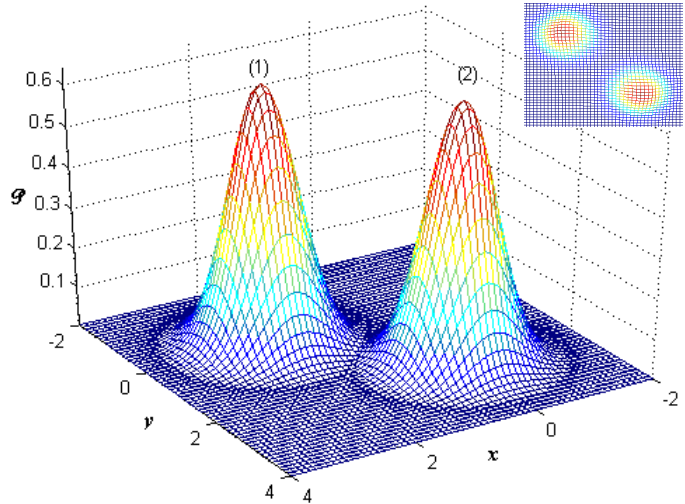

**Fig. 3: Amplitude probabilities distributions for the quantum states $|H\rangle=|2,0\rangle$ (1) and $|V\rangle=|0,2\rangle$ (2).**

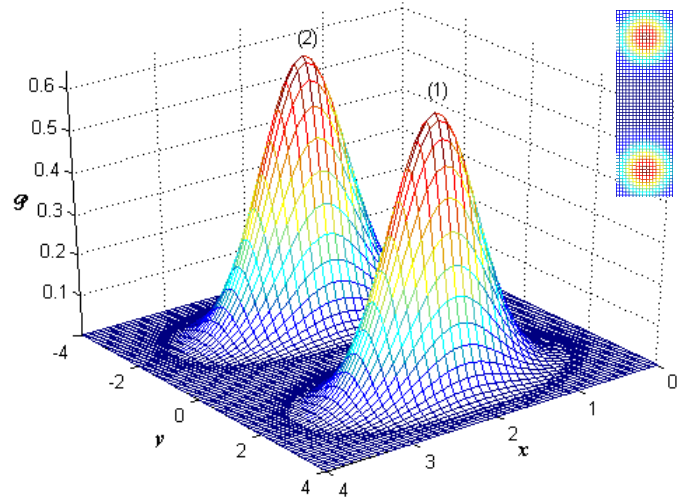

**Fig. 4: Amplitude probabilities distributions for the states $|\pi/4\rangle=|2,2\rangle$ (1) and $|-\pi/4\rangle=|2,-2\rangle$ (2).**

We also obtained numerically $\langle x\rangle\approx 2$ and $\langle y\rangle\approx 0$ for $|H\rangle=|2,0\rangle$ and $\langle x\rangle\approx 0$ and $\langle y\rangle\approx 2$ for $|V\rangle=|0,2\rangle$. Similarly, we obtained $\langle x\rangle\approx 2$ and $\langle y\rangle\approx 2$ for $|\pi/4\rangle=|2,2\rangle$ and $\langle x\rangle\approx 2$ and $\langle y\rangle\approx -2$ for $|-\pi/4\rangle=|2,-2\rangle$. Regarding the circular polarizations, its amplitude probabilities distributions are plotted in Fig. 5 [15,18].

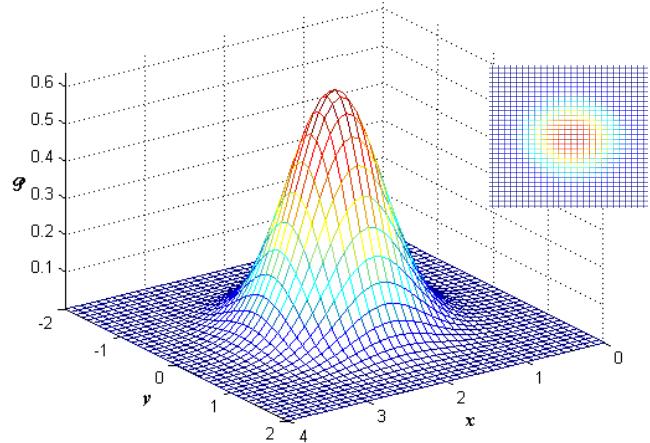

**Fig. 5: Amplitude probabilities distributions for the states $|RC\rangle=|2,2i\rangle$ and $|LC\rangle=|2,-2i\rangle$.**

The mean values obtained numerically for the amplitudes are $\langle x\rangle\approx 2$ and $\langle y\rangle\approx 0$ for both $|RC\rangle=|2,2i\rangle$ and $|LC\rangle=|2,-2i\rangle$.

## 3. Quantum polarization of superposition of two two-mode coherent states

In this section we consider the quantum Stokes parameters and the quantum degree of polarization of quantum states composed by the superposition of two bimodal coherent states

$|\psi\rangle = N(|\alpha,\beta\rangle + |\varepsilon,\lambda\rangle)$, where $|N|^2 = \left\{2 + \left[\zeta + \zeta^*\right]\exp\left[-\left(|\alpha|^2 + |\beta|^2 + |\varepsilon|^2 + |\lambda|^2\right)/2\right]\right\}^{-1}$ and $\zeta = \exp(\alpha^*\varepsilon + \beta^*\lambda)$. The averages of the quantum Stokes parameters and of their squared values, for $|\psi\rangle$, are:

$$\langle \hat{S}_1 \rangle = |N|^2 \left\{ \left(|\alpha|^2 - |\beta|^2\right) + \left(|\varepsilon|^2 - |\lambda|^2\right) + \left[\left(\alpha^*\varepsilon - \beta^*\lambda\right) + \left(\alpha\varepsilon^* - \beta\lambda^*\right)\right]\delta \right\} \quad (22)$$

$$\langle \hat{S}_2 \rangle = |N|^2 \left\{ \left(\alpha^*\beta + \alpha\beta^*\right) + \left(\varepsilon^*\lambda + \varepsilon\lambda^*\right) + \left[\left(\alpha^*\lambda + \varepsilon\beta^*\right) + \left(\varepsilon^*\beta + \alpha\lambda^*\right)\right]\delta \right\} \quad (23)$$

$$\langle \hat{S}_3 \rangle = j|N|^2 \left\{ \left(\alpha\beta^* - \alpha^*\beta\right) + \left(\varepsilon\lambda^* - \varepsilon^*\lambda\right) + \left[\left(\varepsilon\beta^* - \alpha^*\lambda\right) + \left(\alpha\lambda^* - \beta\varepsilon^*\right)\right]\delta \right\} \quad (24)$$

$$\langle \hat{S}_1^2 \rangle = |N|^2 \left\{ \begin{array}{l} |\alpha|^2 + |\beta|^2 + |\varepsilon|^2 + |\lambda|^2 + \left(|\alpha|^2 - |\beta|^2\right)^2 + \left(|\varepsilon|^2 - |\lambda|^2\right)^2 + \\ \left[\alpha^*\varepsilon + \beta^*\lambda + \alpha\varepsilon^* + \beta\lambda^* + \left(\alpha^*\varepsilon - \beta^*\lambda\right)^2 + \left(\alpha\varepsilon^* - \beta\lambda^*\right)^2\right]\delta \end{array} \right\} \quad (25)$$

$$\langle \hat{S}_2^2 \rangle = |N|^2 \left\{ \begin{array}{l} |\alpha|^2 + |\beta|^2 + |\varepsilon|^2 + |\lambda|^2 + 2\left(|\alpha|^2|\beta|^2 + |\varepsilon|^2|\lambda|^2\right) + \left(\alpha^*\beta\right)^2 + \left(\alpha\beta^*\right)^2 + \left(\varepsilon^*\lambda\right)^2 + \\ \left(\varepsilon\lambda^*\right)^2 + \left[\alpha^*\varepsilon + \beta^*\lambda + \alpha\varepsilon^* + \beta\lambda^* + \left(\alpha^*\lambda + \beta^*\varepsilon\right)^2 + \left(\varepsilon^*\beta + \alpha\lambda^*\right)^2\right]\delta \end{array} \right\} \quad (26)$$

$$\langle \hat{S}_3^2 \rangle = |N|^2 \left\{ \begin{array}{l} |\alpha|^2 + |\beta|^2 + |\varepsilon|^2 + |\lambda|^2 + 2\left(|\alpha|^2|\beta|^2 + |\varepsilon|^2|\lambda|^2\right) - \left(\alpha^*\beta\right)^2 - \left(\alpha\beta^*\right)^2 - \left(\varepsilon^*\lambda\right)^2 - \\ \left(\varepsilon\lambda^*\right)^2 + \left[\alpha^*\varepsilon + \beta^*\lambda + \alpha\varepsilon^* + \beta\lambda^* - \left(\alpha^*\lambda - \beta^*\varepsilon\right)^2 - \left(\varepsilon^*\beta - \alpha\lambda^*\right)^2\right]\delta \end{array} \right\} \quad (27)$$

where $\delta = \exp\left[\alpha^*\varepsilon + \beta^*\lambda - \left(|\alpha|^2 + |\beta|^2 + |\varepsilon|^2 + |\lambda|^2\right)/2\right]$. On the other hand, the $Q$ function of the state $|\psi\rangle = N(|\alpha,\beta\rangle + |\varepsilon,\lambda\rangle)$ is given by

$$Q(\theta,\phi) = \frac{|N|^2}{4\pi} \left\{ \begin{array}{l} e^{-\left(|\alpha|^2 + |\beta|^2\right)}(1+z_1)e^{z_1} + e^{-\left(|\varepsilon|^2 + |\lambda|^2\right)}(1+z_2)e^{z_2} + \\ e^{-\left(|\alpha|^2 + |\beta|^2 + |\varepsilon|^2 + |\lambda|^2\right)/2}\left[(1+z_{12})e^{z_{12}} + (1+z_{12}^*)e^{z_{12}^*}\right] \end{array} \right\}, \quad (28)$$

$$z_1 = \left[|\alpha|\cos\left(\frac{\theta}{2}\right)\cos(\phi_\alpha + \phi) + |\beta|\text{sen}\left(\frac{\theta}{2}\right)\cos(\phi_\beta)\right]^2 + \left[|\alpha|\cos\left(\frac{\theta}{2}\right)\text{sen}(\phi_\alpha + \phi) + |\beta|\text{sen}\left(\frac{\theta}{2}\right)\text{sen}(\phi_\beta)\right]^2, \quad (29)$$

$$z_2 = \left[|\varepsilon|\cos\left(\frac{\theta}{2}\right)\cos(\phi_\varepsilon + \phi) + |\lambda|\text{sen}\left(\frac{\theta}{2}\right)\cos(\phi_\lambda)\right]^2 + \left[|\varepsilon|\cos\left(\frac{\theta}{2}\right)\text{sen}(\phi_\varepsilon + \phi) + |\lambda|\text{sen}\left(\frac{\theta}{2}\right)\text{sen}(\phi_\lambda)\right]^2, \quad (30)$$

$$z_{12} = |\alpha||\varepsilon|\cos^2\left(\frac{\theta}{2}\right)e^{j(\phi_\alpha - \phi_\varepsilon)} + |\beta||\lambda|\text{sen}^2\left(\frac{\theta}{2}\right)e^{j(\phi_\beta - \phi_\lambda)} + \left[\frac{|\alpha||\lambda|e^{j(\phi_\alpha - \phi_\lambda + \phi)} + |\beta||\varepsilon|e^{j(\phi_\beta - \phi_\varepsilon - \phi)}}{2}\right]\text{sen}(\theta). \quad (31)$$

In this work, we are going to be concerned only with the following particular cases of $|\psi\rangle$: $|\psi_1\rangle=N_1(|\alpha,\beta\rangle+|\beta,\alpha\rangle)$, $|\psi_2\rangle=N_2(|-\alpha,-\alpha\rangle+|\alpha,\alpha\rangle)$ and $|\psi_3\rangle=N_3(|\alpha,0\rangle+|0,\alpha\rangle)$, where $N_1=[2(1+e^{2\alpha\beta-(|\alpha|^2+|\beta|^2)})]^{-1/2}$, $N_2=[2(1+e^{-4|\alpha|^2})]^{-1/2}$ and $N_3=[2(1+e^{-|\alpha|^2})]^{-1/2}$, for $\alpha$ and $\beta$ real values. From (22)-(27), the averages and variances of the quantum Stokes parameters of the states $|\psi_1\rangle$, $|\psi_2\rangle$ and $|\psi_3\rangle$ are:

$$|\psi_1\rangle \begin{cases} \langle \hat{S}_1 \rangle = 0, & V_1 = 2N_1^2\left[\alpha^2+\beta^2+2\alpha\beta\delta_1+\left(\alpha^2-\beta^2\right)^2\right], \\ \langle \hat{S}_2 \rangle = 2N_1^2\left[2\alpha\beta+\left(\alpha^2+\beta^2\right)\delta_1\right], & V_2 = 2N_1^2\left[\begin{array}{l}\left\{+2\alpha\beta(2\alpha\beta+\delta_1)+\left(\alpha^2+\beta^2\right)\left[1+\left(\alpha^2+\beta^2\right)\delta_1\right]\right\}-\\ 2N_1^2\left[2\alpha\beta+\left(\alpha^2+\beta^2\right)\delta_1\right]^2\end{array}\right], \\ \langle \hat{S}_3 \rangle = 0, & V_3 = 2N_1^2\left[4\alpha\beta(1+\alpha\beta)-\left(\alpha^4+\beta^4\right)\delta_1\right], \\ \delta_1 = e^{\left[2\alpha\beta-\left(|\alpha|^2+|\beta|^2\right)\right]}; \end{cases} \quad (32)$$

$$|\psi_2\rangle \begin{cases} \langle \hat{S}_1 \rangle = 0, & V_1 = 4N_2^2\alpha^2(1-\delta_2), \\ \langle \hat{S}_2 \rangle = 4N_2^2\alpha^2(1+\delta_2), & V_2 = 4N_2^2\alpha^2\left\{\left[(1+2\alpha^2)-(1-2\alpha^2)\delta_2\right]-4N_2^2\alpha^2(1+\delta_2)^2\right\}, \\ \langle \hat{S}_3 \rangle = 0, & V_3 = 4N_2^2\alpha^2(1-\delta_2), \\ \delta_2 = e^{-4\alpha^2}; \end{cases} \quad (33)$$

$$|\psi_3\rangle \begin{cases} \langle \hat{S}_1 \rangle = 0, & V_1 = 2N_3^2\alpha^2(1+\alpha^2), \\ \langle \hat{S}_2 \rangle = 2N_3^2\alpha^2\delta_3, & V_2 = 2N_3^2\alpha^2\left[1+\alpha^2(1-2N_3^2\delta_3)\delta_3\right], \\ \langle \hat{S}_3 \rangle = 0, & V_3 = 2N_3^2\alpha^2(1-\alpha^2\delta_3), \\ \delta_3 = e^{-\alpha^2}. \end{cases} \quad (34)$$

One can observe in (32)-(34) that $\langle S_1 \rangle$ and $\langle S_3 \rangle$ are equal to zero for the three states. This happens because the average optical powers in horizontal and vertical polarizations are equals, for all of them. Hence, $|\psi_1\rangle$, $|\psi_2\rangle$ and $|\psi_3\rangle$ correspond to the polarization state $|\pi/4\rangle$. An interesting behavior caused by the superposition can be seen at the variances of the Stokes parameters. For states of the type $|\alpha,\beta\rangle$, the variances of all Stokes parameters are equal and they vary proportionally to the optical power, as can be seen in (7)-(9). However, for states of the type $|\psi_1\rangle=N_1(|\alpha,\beta\rangle+|\beta,\alpha\rangle)$, the variances of the Stokes parameters can increase and decrease when the total mean photon number increases. This can be seen in Fig. 6 for $V_3$.

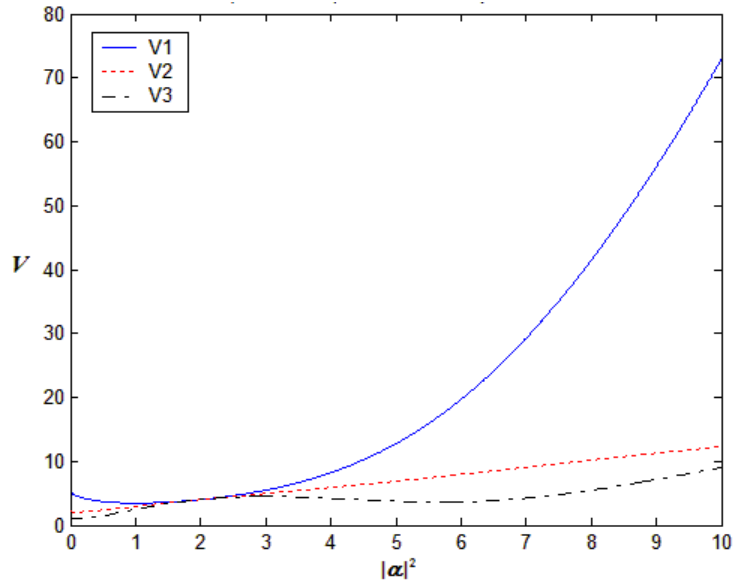

**Figure 6:** Variances of $S_1$ ($V_1$), $S_2$ ($V_2$) and $S_3$ ($V_3$), versus $|\alpha|^2$, for $|\psi_1\rangle$ having $|\beta|^2=4$.

Now, using (28)-(31) we obtain the following $Q$ functions for $|\psi_1\rangle$, $|\psi_2\rangle$ and $|\psi_3\rangle$:

$$|\psi\rangle_1 \begin{cases} Q_1(\theta,\phi) = \dfrac{e^{-(|\alpha|^2+|\beta|^2)}|N_1|^2}{4\pi}\{(1+z_1)e^{z_1}+(1+z_2)e^{z_2}+2(1+z_{12})e^{z_{12}}\}, \\ z_1 = |\alpha|^2 \cos^2\left(\dfrac{\theta}{2}\right)+|\beta|^2 \operatorname{sen}^2\left(\dfrac{\theta}{2}\right)+|\alpha||\beta|\operatorname{sen}(\theta)\cos(\phi), \\ z_2 = |\beta|^2 \cos^2\left(\dfrac{\theta}{2}\right)+|\alpha|^2 \operatorname{sen}^2\left(\dfrac{\theta}{2}\right)+|\alpha||\beta|\operatorname{sen}(\theta)\cos(\phi), \\ z_{12} = |\alpha||\beta|[1+\operatorname{sen}(\theta)\cos(\phi)]; \end{cases} \quad (35)$$

$$|\psi\rangle_2 \begin{cases} Q_2(\theta,\phi) = \dfrac{e^{-2|\alpha|^2}|N_2|^2}{2\pi}\{(1+z_1)e^{z_1}+(1+z_{12})e^{z_{12}}\}, \\ z_1 = |\alpha|^2[1+\operatorname{sen}(\theta)\cos(\phi)], \quad z_{12} = -|\alpha|^2[1+\operatorname{sen}(\theta)\cos(\phi)]; \end{cases} \quad (36)$$

$$|\psi\rangle_3 \begin{cases} Q_3(\theta,\phi) = \dfrac{e^{-|\alpha|^2}|N_3|^2}{4\pi}\{(1+z_1)e^{z_1}+(1+z_2)e^{z_2}+2(1+z_{12})e^{z_{12}}\}, \\ z_1 = |\alpha|^2 \cos^2\left(\dfrac{\theta}{2}\right), \quad z_2 = |\alpha|^2 \operatorname{sen}^2\left(\dfrac{\theta}{2}\right), \quad z_{12} = |\alpha|^2 \dfrac{e^{j\phi}}{2}\operatorname{sen}(\theta). \end{cases} \quad (37)$$

As an illustration, using (35)-(37), one can obtain the $Q$ functions of the states $|\psi_1\rangle$, $|\psi_2\rangle$ and $|\psi_3\rangle$, having $|\alpha|^2=|\beta|^2=4$. The results are shown in Fig. 7.

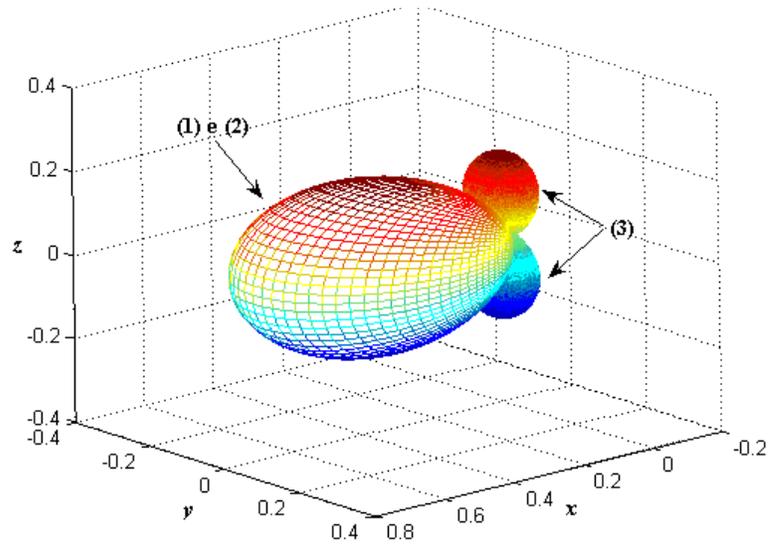

**Figure 7:** $Q$ functions of the states $|\psi_1\rangle$ (1), $|\psi_2\rangle$ (2) and $|\psi_3\rangle$ (3) obtained using (35)-(37) having $|\alpha|^2 = |\beta|^2 = 4$.

Finally, using (35)-(37) in (15)-(16), we calculated the quantum degree of polarization of the states $|\psi_1\rangle$ ($|\beta|^2$=0.5, 1, 2, 4), $|\psi_2\rangle$ and $|\psi_3\rangle$ versus $|\alpha|^2$. The results are shown in Fig. 8.

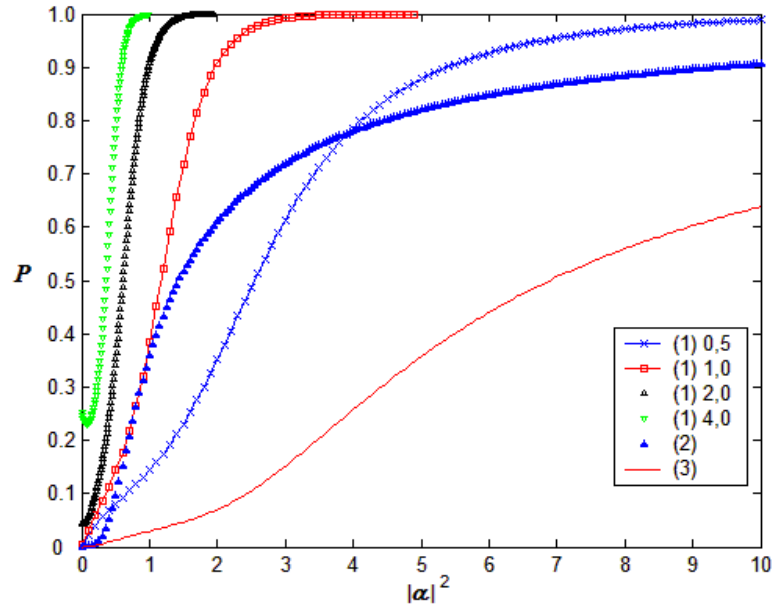

**Fig. 8:** Quantum degree of polarization, $P$, versus $|\alpha|^2$ for states $|\psi_1\rangle$(1), $|\psi_2\rangle$ (2) and $|\psi_3\rangle$(3).

From Fig. 8, is straightforward to note that, the larger the mean photon number the larger is the quantum degree of polarization.

## 4. Quantum entanglement of superposition of two two-mode coherent states

The quantum entanglement of the state $|\psi\rangle = N(|\alpha,\beta\rangle + |\varepsilon,\lambda\rangle)$ can be measured by the concurrence described in [19,20]

$$C = \frac{\sqrt{(1-|\langle\alpha|\varepsilon\rangle|^2)(1-|\langle\lambda|\beta\rangle|^2)}}{1+\text{Re}\{\langle\alpha|\varepsilon\rangle\langle\beta|\lambda\rangle\}} \tag{38}$$

The concurrences of $|\psi_1\rangle$, $|\psi_2\rangle$ and $|\psi_3\rangle$ are, respectively:

$$C_1 = \frac{1-e^{-|\alpha-\beta|^2}}{1+e^{-|\alpha-\beta|^2}}, \quad C_2 = \frac{1-e^{-4|\alpha|^2}}{1+e^{-4|\alpha|^2}}, \quad C_3 = \frac{1-e^{-|\alpha|^2}}{1+e^{-|\alpha|^2}}. \tag{39}$$

The plot of $C_1$ versus $|\alpha|^2$ and $|\beta|^2$ can be seen in Fig. 9.

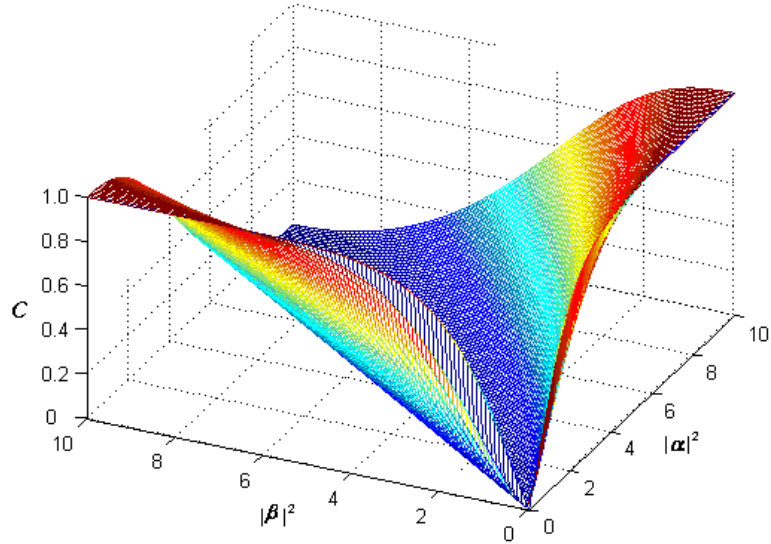

**Fig. 9: Concurrence of the state $|\psi_1\rangle = N_1(|\alpha,\beta\rangle + |\beta,\alpha\rangle)$ versus $|\alpha|^2$ and $|\beta|^2$.**

From (39) we see there will exist entanglement in $|\psi_1\rangle$ only if $\alpha \neq \beta$. It is also easy to see from $C_2$ and $C_3$ equations that, the larger the value of $|\alpha|^2$ the large are the entanglements of $|\psi_2\rangle$ and $|\psi_3\rangle$.

Now, let us assume the state $|\psi_1\rangle$ pass by a compensator-rotacionator-compensator device, $C(\phi_2)R(\theta)C(\phi_1)$. The quantum state at the output and its entanglement are given by

$$|\psi_4\rangle = N_1 \begin{pmatrix} \left|\beta\mathrm{sen}(\theta)e^{j(\phi_2-\phi_1)/2}+\alpha e^{j(\phi_2+\phi_1)/2}\cos(\theta)\right\rangle \left|\beta e^{-j(\phi_2+\phi_1)/2}\cos(\theta)-\alpha e^{-j(\phi_2-\phi_1)/2}\mathrm{sen}(\theta)\right\rangle + \\ \left|\alpha\mathrm{sen}(\theta)e^{j(\phi_2-\phi_1)/2}+\beta e^{j(\phi_2+\phi_1)/2}\cos(\theta)\right\rangle \left|\alpha e^{-j(\phi_2+\phi_1)/2}\cos(\theta)-\beta e^{-j(\phi_2-\phi_1)/2}\mathrm{sen}(\theta)\right\rangle \end{pmatrix}, \quad (40)$$

$$C_4 = \frac{\sqrt{1+e^{-2|\alpha-\beta|^2}-2e^{-|\alpha-\beta|^2}\cosh\left[|\alpha-\beta|^2\mathrm{sen}(2\theta)\cos(\phi_1)\right]}}{1+e^{-|\alpha-\beta|^2}}. \quad (41)$$

Figure 10 shows the change in the entanglement according to variations in $\phi_1$ and $\theta$, for $|\alpha-\beta|^2=4$.

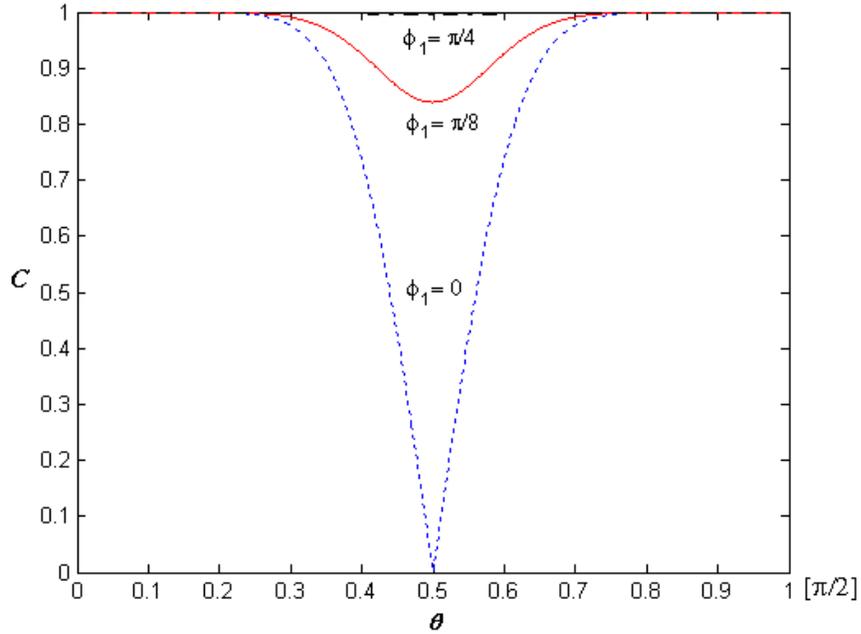

**Figure 10:** Concurrence of $|\psi_4\rangle$(Eq. (40)) versus $\theta$ (rotator's angle) for $\phi_1 \in \{0, \pi/8, \pi/4\}$ and $|\alpha-\beta|^2=4$.

It is interesting the fact that the rotation of $\pi/4$ applied in the state $N_1(|\alpha,\beta\rangle+|\beta,\alpha\rangle)$ destroys completely its entanglement: $R(\pi/4)N_1(|\alpha,\beta\rangle+|\beta,\alpha\rangle) = N_1(|(\alpha-\beta)/2^{1/2}\rangle+|(\beta-\alpha)/2^{1/2}\rangle)|(\beta+\alpha)/2^{1/2}\rangle$. This implies the polarization rotator $R(\pi/4)$ can be a perfect entangler/disentangler gate.

## 5. Conclusions

We have presented explicit formulas for the mean values and the variances of the quantum Stokes parameters, as well the $Q$ functions of superposition of two-mode coherent states. The results showed that the superposition has a great influence in variances of the Stokes parameters. The decrease of the variance of $S_3$ of the state $|\psi_1\rangle$ when the mean photon number increases, shows that the uncertainty about polarization can decrease when optical power increases. This does not happen when normal two-mode coherent states are considered. Moreover, for the cases studied, the degree of polarization still increases monotonically with the total mean photon number, as happens with normal two-mode coherent states. At last, we showed that the polarization rotator can change the quantum entanglement of the superposition of two-mode coherent states, in particular, the polarization rotator with angle equal to $\pi/4$ can be a perfect entangler/disentangler gate.

# References

[1] A. Vidiella-Barranco and L.F.M. Borelli, Int. J. Mod. Phys. B., 20, 1287 (2006).

[2] G. A. Barbosa, Phys. Rev. A, 68, 052307 (2003).

[3] Zhen-Qiang Yin, Zheng-Fu Han, Fang-Wen Sun and Guang-Can Guo, Phys. Rev. A, 76, 014304 (2007).

[4] Won-Ho Kye, Chil-Min Kim, M. S. Kim, and Young-Jai Park, Phys. Rev. Lett., 95, 040501 (2005).

[5] J. B. R. Silva and R. V. Ramos, Optics Comm., vol. 281, no. 9, p. 2705-2709 (2008).

[6] S. Glancy, H. Vasconcelos and T. C. Ralph, Phys. Rev. A, 70, 22317 (2004).

[7] H. Jeong and M. Kim, Phys. Rev. A, 65, 042305 (2002).

[8] H. H. M. de Vasconcelos. *Topics in coherent state quantum computation and state purification*. Doctor of philosophy thesis, Graduate Program in Physics Notre Dame, Indiana (2006).

[9] B.A. Robson, The Theory of Polarization Phenomena (Clarendon, Oxford, 1974).

[10] A.S. Chirkin, A.A. Orlov and D. Yu. Paraschuk, Quant. Electron. 23 870 (1993).

[11] P. Usachev, J. Söderholm, G. Björk, et al., Optics Commun. 193 161 (2001).

[12] G.S. Agarwal, J. Lehner and H. Paul, Optics Commun. 129 369 (1996).

[13] H. Prakash and N. Chandra, Phys. Rev. A 4 796 (1971).

[14] J. Lehner, U. Leonhardt and H. Paul, Phys. Rev. A 53 2727 (1996).

[15] A. Luis, "Degree of polarization in quantum optics", Phys. Rev. A, 66, 013806 (2002).

[16] A. B. Klimov, L. L. Sánchez-Soto, E. C. Yustas, J. Söderholm and G. Björk, quant-ph/0504226 (2005).

[17] R. V. Ramos, J. of Mod. Opt., Vol. 52, No. 15, 2093–2103 (2005).

[18] T. Tsegaye, J. Söderholm, Mete Atatüre, Alexei Trifonov, Gunnar Björk, Alexander V. Sergienko, Bahaa E. A. Saleh and Malvin C. Teich, Phys. Rev. Lett. 85, 5013 (2000).

[19] P. Rungta, V. Bužek, C. M. Caves, M. Hillery and G. J. Milburn, Phys. Rev. A 64, 042315 (2001).

[20] L. M. Kuang and L. Zhou, Phys. Rev. A 68, 043606 (2003).

[21] A. Ourjoumtsev, R. Tualle-Brouri, J. Laurat, and P. Grangier, Science 312, 83-86 (2006).

[22] S. Glancy and H. M. Vasconcelos, arXiv:0705.2045 (2007).